# Z Theory

## and its
## Quantum-Relativistic Operators

Pietro Giorgio Zerbo

(with the kind contribution of Alessandro Alemberti)

# Abstract

*The view provided by Z theory, based on its quantum-relativistic operators, is an integrated picture of the micro and macro quantities relationships.*
*The axiomatic formulation of the theory is presented in this paper.*
*The theory starts with the existence of the wave function, the existence of three fundamental constants h, c and G as well as the physical quantity $R_c$ (the radius of the space-time continuum) plus the definition of a general form for the quantum-relativistic functional operators.*
*Using such starting point the relationships between relativity, quantum mechanics and cosmological quantities can be clarified.*

# Index

**Basic notation**

As a first step we need to clarify the notation used in this work. The space-time co-ordinates are expressed as follows:

$$\boldsymbol{s} = (ct, i\vec{x}) = (ct, ix, iy, iz) = (s_1, s_2, s_3, s_4)$$

where $i$ is the imaginary unit.

Consistently, the momentum four-vector is:

$$\boldsymbol{p} = (\frac{E}{c}, i\vec{p}) = (\frac{E}{c}, ip_x, ip_y, ip_z) = (p_1, p_2, p_3, p_4)$$

We form the dot product with itself of each of the two four-vectors above to obtain:

$$\boldsymbol{s} \circ \boldsymbol{s} = c^2 t^2 - x^2 - y^2 - z^2 = s^2$$

$$\boldsymbol{p} \circ \boldsymbol{p} = E^2/c^2 - p_x^2 - p_y^2 - p_z^2 = m_0^2 c^2$$

to realise that such dot products are equal to the relativistic space-time continuum equation and to the relativistic energy equation, respectively.

Only using the above four-vector representation we achieve a real number for the rest mass $m_0$ and for the space-time interval $s$ when the particle speed is less then the speed of light.

On the basis of such a definition of the four-vector quantities we define the four-gradient as follows:

$$\frac{\partial}{\partial s_n}\bigg|_{n=1,4} = (\frac{\partial}{c\partial t}, \frac{\partial}{i\partial x}, \frac{\partial}{i\partial y}, \frac{\partial}{i\partial z})$$

This notation will be used throughout.

## 1. Z Theory: the axiomatic formulation

The axiomatic formulation of Z theory is based on the following set of fundamental statements:

# Axioms

1) The state of a physical system is described by its own wave function $\Psi$.

2) There are three fundamental constants in the physical world: the Planck constant $h$, the speed of light $c$ and the gravitational constant $G$. In addition there is a fundamental quantity: the radius of the space-time continuum, $R_c$.

3) The general form of the quantum-relativistic functional operators is:

$$[\mathbf{QO}] = \frac{K_{QO}}{2\pi i} \frac{\partial}{\partial s_n}\bigg|_{n=1,4} \qquad [\mathbf{QO}] = \frac{K_{QO}}{2\pi i}\left(\frac{\partial}{c\partial t}, \frac{\partial}{i\partial x}, \frac{\partial}{i\partial y}, \frac{\partial}{i\partial z}\right)$$

where $K_{QO}$ is to be substituted with the three constants $h$, $c$, $G$ or $R_c^2$.

4) Application of each quantum-relativistic functional operator to the wave function $\Psi$ generates a physical quantity which is the physical aspect of the operator. Additional physical quantities arise using different combinations of the four basic functional operators.

Our work begins from this set of axioms using the powerful tool of the quantum-relativistic functional operators as defined previously.

However, before presenting the analysis we need to derive the general form of the four-vectors and of the algebraic equations associated to four-vectors in detail. This is done in the following steps.

**General form of the four-vectors**

The third axiom contains the general form of the four-operator, which will not be repeated here. To explicit the structure of the four-vectors we first need to define the wave function. For the free particle we will use:

$$\Psi = e^{\frac{2\pi i}{h}(Et - \vec{p} \circ \vec{x})} \quad \text{and this means that:} \quad \Psi = e^{\frac{2\pi i}{h}(\boldsymbol{p} \circ \boldsymbol{s})}$$

The latter expression of the wave function shows its relativistic invariance immediately, the exponent being the result of the dot product of the two position and momentum four-vectors multiplied by constants.

The general form of the quantum-relativistic operator is applied to the wave function of the free particle to obtain:

$$\frac{K_{QO}}{2\pi i} \frac{\partial \Psi}{c \partial t} = \frac{K_{QO}}{hc} E \Psi \qquad \frac{K_{QO}}{2\pi i} \frac{\partial \Psi}{i \partial x_\alpha} = -\frac{K_{QO}}{ih} p_\alpha \Psi \quad \text{where: } \alpha = 1, 2, 3$$

we can use $E = mc^2$ in the spatial component expressions to get:

$$\frac{K_{QO}}{2\pi i} \frac{\partial \Psi}{i \partial x_\alpha} = \frac{i K_{QO}}{h} \frac{E}{c^2} v_\alpha \Psi$$

so that: $[QO] \Psi = \dfrac{K_{QO}}{hc} E (1, i\beta_x, i\beta_y, i\beta_z) \Psi$ where: $\beta_\alpha = \dfrac{v_\alpha}{c}$.

This is the general form of the correspondence between the four-operator **[QO]** and the associated four-vector:

$$\boldsymbol{qv} = \frac{K_{QO}}{hc} E (1, i\beta_x, i\beta_y, i\beta_z), \text{ in other words:}$$

$$[QO] \Psi = \boldsymbol{qv} \Psi$$

which states that $\boldsymbol{qv}$ is an eigen-vector of the four-operator **[QO]**.

**Algebraic and differential equations general form**

We now use the expression of the *qv* four-vector in the dot product with itself:

$$\mathbf{qv} \circ \mathbf{qv} = \frac{K_{QO}^2}{h^2 c^2} E^2 (1 - \beta^2) = q_0^2$$

This quantity is a relativistic invariant, in particular equal to its value in the reference system of the particle, $q_0^2$. Consequently, we can write:

$$\frac{K_{QO}^2}{h^2 c^2} E^2 (1 - \beta^2) = \frac{K_{QO}^2}{h^2 c^2} E_0^2$$

This is the general form of the algebraic equations associated to the dot product of each four-vector with itself.
Using the same process, we can derive the general form of the differential equations associated to the dot product of each four-operator with itself. Applying the four-operator dot product to the wave function, we have:

$$[\mathbf{QO}][\mathbf{QO}]\,\Psi = [\mathbf{QO}]^2\,\Psi = \frac{K_{QO}^2}{4\pi^2} \left( \frac{\partial^2 \Psi}{\partial x^2} + \frac{\partial^2 \Psi}{\partial y^2} + \frac{\partial^2 \Psi}{\partial z^2} - \frac{1}{c^2} \frac{\partial^2 \Psi}{\partial t^2} \right)$$

and we may write this equation in a more compact form using the d'Alembert operator:

$$\Box^2 \Psi = \frac{4\pi^2}{K_{QO}^2} [\mathbf{QO}]^2 \Psi$$

This is the expression of the general differential equations obtained from the definition of the four-operators. It has to be noted that the d'Alembert operator is relativistically invariant. As a consequence, when applied to the wave function, it produces values which are equal to that calculated in the reference system of the considered phenomena. Thus:

$$\Box^2 \Psi = \frac{4\pi^2}{K_{QO}^2} q_0^2 \Psi$$

where the quantity $q_0$ on the right side can be easily identified using considerations of dimensional analysis based on the dimensions of each different quantity $K_{QO}$.

## 2. Differential equations and equivalent physical descriptions

On the basis of the above considerations the differential operators defined through the use of the three fundamental constants and of the evolutionary variable generate a system of four differential equations:

$h$ : momentum
$$\frac{\partial^2 \Psi}{\partial x^2} + \frac{\partial^2 \Psi}{\partial y^2} + \frac{\partial^2 \Psi}{\partial z^2} - \frac{1}{c^2}\frac{\partial^2 \Psi}{\partial t^2} = \frac{4\pi^2 p_0^2}{h^2} \Psi$$

$c$ : frequency
$$\frac{\partial^2 \Psi}{\partial x^2} + \frac{\partial^2 \Psi}{\partial y^2} + \frac{\partial^2 \Psi}{\partial z^2} - \frac{1}{c^2}\frac{\partial^2 \Psi}{\partial t^2} = \frac{4\pi^2 \nu_0^2}{c^2} \Psi$$

$G$: gravitation
$$\frac{\partial^2 \Psi}{\partial x^2} + \frac{\partial^2 \Psi}{\partial y^2} + \frac{\partial^2 \Psi}{\partial z^2} - \frac{1}{c^2}\frac{\partial^2 \Psi}{\partial t^2} = \frac{4\pi^2 g_0^2}{G^2} \Psi$$

$R_c^2$: position
$$\frac{\partial^2 \Psi}{\partial x^2} + \frac{\partial^2 \Psi}{\partial y^2} + \frac{\partial^2 \Psi}{\partial z^2} - \frac{1}{c^2}\frac{\partial^2 \Psi}{\partial t^2} = \frac{4\pi^2 s_0^2}{R_c^4} \Psi$$

The first equation, obtained from the momentum four-operator, is recognised as the Klein-Gordon equation for a relativistic free particle, when we substitute the quantity $4\pi^2 p_0^2/h^2$ with $m_0^2 c^2/\hbar^2$.

The second is a wave equation which describes a vibrational phenomenon having a frequency $\nu_0$. The third equation concerns the gravitational aspect of the physical reality. We need to go into more detail but, for the moment, we just observe that $g_0$ has the dimensions of a squared velocity divided by a mass. The last equation deals with the space-time aspect of the physical reality. It is sufficient here to note that the quantity $s_0$ has the dimension of a length and, actually, is a relativistic space-time interval.

The main property of the above differential equations system is that the left hand side members of the differential equations are all equal to the wave equation. On the basis of this property we need to point out the following very important fact: **all equations may be considered as different aspects of the same physical reality**.
In other words talking about a particle proper mass has the same meaning as talking about its proper frequency, its proper gravitation or its proper space-time interval. This fact will allow us to deduce a very important set of equivalence laws, some of them already known and verified experimentally, some of them completely new. We are now at the heart of Z theory and we are going to explain the very important equivalence relationships in the following steps.

**Momentum and frequency**

$$\frac{4\pi^2 \nu_0^2}{c^2} = \frac{4\pi^2 m_0^2 c^2}{h^2} \rightarrow \quad h^2 \nu_0^2 = m_0^2 c^4 \rightarrow \quad \left| h\nu_0 \right| = \left| m_0 c^2 \right|$$

This first equivalence between the eigen values of the momentum and frequency functional operators generates, from the axioms of Z theory, the equivalence law discovered by L.V. De Broglie.
We have to point out that the form here deduced is more general since it involves the squared values of frequency and mass. This fact is obviously true for all equivalency laws we will analyse in the next steps. However, for the sake of simplicity, it will be omitted in the following presentation.

**Momentum and gravitation**

$$\frac{4\pi^2 g_0^2}{G^2} = \frac{4\pi^2 m_0^2 c^2}{h^2}$$

as we have already noted, $g_0$ has the dimensions of a squared velocity divided by a mass, so if we define:

$$g_0 = \frac{c^2}{M_0} \quad \text{we obtain:} \quad m_0 M_0 = \frac{hc}{G} = M_p^2$$

In other words, the definition of the Planck mass arises in a natural way when one compares the momentum and gravitational description of the considered phenomena.
Let's look at the formulation of a new concept: since the quantity $M_0$ is strictly related to the particle of mass $m_0$ through the above hyperbolic relationship, we will call $M_0$ the proper mass of the eigen-universe associated to the particle.

**Momentum and position**

$$\frac{4\pi^2 m_0^2 c^2}{h^2} = \frac{4\pi^2 s_0^2}{R_c^4} \rightarrow m_0^2 = \frac{h^2 s_0^2}{c^2 R_c^4} \rightarrow m_0 = \frac{h}{c R_c^2} s_0$$

We are going to analyse in more detail this relationship for the following two particular cases:

$$s_0 = 0 \quad \rightarrow \quad \frac{m_0^2 c^2}{h^2} = 0 \quad \rightarrow \quad m_0^2 = 0 \quad \rightarrow \quad m_0 = 0$$

If the space-time interval is null the mass of the correspondent particle is equal to zero.

$$s_0 = R_c \quad \rightarrow \quad \frac{m_0^2 c^2}{h^2} = \frac{1}{R_c^2} \quad \rightarrow \quad m_0^2 = \frac{h^2}{c^2 R_c^2} \quad \rightarrow \quad m_0 = \frac{h}{c R_c}$$

If the space-time interval is equal to the radius of our space-time continuum ($R_c = ct_c = ct_{UNI}$), the mass of the correspondent particle will be equal to $h/cR_c$. It is quite obvious now to observe that this relationship, expressed as: $R_c = h/m_0 c$, is the usual definition of the particle Compton wavelength:

$$\lambda_c = \frac{h}{m_0 c}$$

As a consequence we can generalise the interpretation of the Compton wavelength concept as the radius of the space-time continuum which is correspondent to a specific particle.
In fact, in the general case, the Compton wavelength of a particle is not necessarily equal to the radius of our space-time continuum.
This brings us to the formulation of two new physical concepts:

### Eigen-universe

The eigen-universe of a particle of proper mass $m_0$ is the space-time continuum with radius $R_0$ such that:

$$R_0 = \frac{h}{m_0 c}$$

### Eigen-particle

The eigen-particle of a universe with radius $R_0$ of its space-time continuum is a particle of proper mass $m_0$ such that:

$$m_0 = \frac{h}{R_0 c}$$

We may now try, with the help of these new concepts, to clarify the equivalence law between momentum and position:

$$\frac{R_c^2}{s_0} = \frac{h}{m_0 c}$$

This expression states that the space-time interval $s_0$ is equivalent to the particle of proper mass $m_0$. As a result we may talk of $s_0$ as the quantum space-time interval correspondent to a particle of proper mass $m_0$. Then we may write the above expression as follows:

$$s_0 R_0 = R_c^2$$

The quantum space-time interval of a particle multiplied by the radius of its eigen-universe is always equal to the square of the radius of our space-time continuum.

**Frequency and gravitation**

$$\frac{4\pi^2 g_0^2}{G^2} = \frac{4\pi^2 \nu_0^2}{c^2} \rightarrow c^2 g_0^2 = \nu_0^2 G^2 \rightarrow c g_0 = \nu_0 G$$

If we use: $g_0 = \dfrac{c^2}{M_0}$ we have: $M_0 = \dfrac{c^3}{G} \dfrac{1}{\nu_0}$ or: $\nu_0 M_0 G = c^3$

which expresses the existing relationship between the mass of a eigen-universe and the correspondent eigen-particle proper frequency. We note, however, that the two non-constant quantities in the above relationship correlate a property of the eigen-universe (mass) to a property of the particle (frequency). We can derive a different equation which is able to involve congruent quantities: both from the universes or both from the particles. In the case of a particle, we will find again the first equivalence law of L.V. De Broglie. To deduce a new equation in the case of a universe, we can apply the equivalence between mass and frequency to an eigen-universe as follows:

$$M_0 c^2 = h N_0$$

So that:
$$M_p^2 c^2 = h N_0 m_0 \quad \rightarrow \quad N_0 m_0 G = c^3$$

Summarising, we can write down the following four relationships covering all the possible combinations between the eigen-universe and eigen-particle masses and frequencies:

$$m_0 c^2 = h \nu_0 \qquad M_0 c^2 = h N_0$$

$$\nu_0 M_0 G = c^3 \qquad N_0 m_0 G = c^3$$

**Frequency and position**

$$\frac{4\pi^2 \nu_0^2}{c^2} = \frac{4\pi^2 s_0^2}{R_c^4} \quad \rightarrow \quad \nu_0^2 = \frac{c^2 s_0^2}{R_c^4} \quad \rightarrow \quad \nu_0 = \frac{c s_0}{R_c^2}$$

Also for this relationship we can consider two different particular cases:

$s_0 = 0 \rightarrow \nu_0 = 0$

if the quantum space-time interval is null the correspondent frequency is equal to zero.

$s_0 = R_c \rightarrow \nu_0 = \dfrac{c}{R_c}$

if the quantum space-time interval is equal to the radius of our space-time continuum (which may be identified with $R_{UNI}$, the radius of our universe) the correspondent frequency is equal to the inverse of our universe age. If this is the case, **such frequency should be identical to the Hubble constant $H_0$.** Moreover, we can say that this frequency will be the proper frequency of the eigen-particle of our universe, i.e. of the graviton, so that:

$$\nu_0 = \nu_H = H_0 = \nu_g$$

Besides the above specific considerations, we will also be able to deduce another very important equation using: $\nu_0 R_c^2 = c s_0$.

If we pose: $R_c = c\, t_c$ and $s_0 = c \tau_0$ (where $\tau_0$ is defined as the quantum proper time of the particle), we obtain: $\nu_0\, t_c^2 = \tau_0$ and taking into account that we can express the oscillation frequency as the inverse of the oscillation period (i.e. $\nu_0 = 1/T_0$), we may write another hyperbolic equation:

$$\tau_0 T_0 = t_c^2$$

In other words the quantum proper time multiplied by the intrinsic oscillation period of a physical entity is always equal to $t_c^2$ which may be identified with the square of the age of our universe.

But we can still develop some additional considerations.
Multiplying both members of $\nu_0 R_c^2 = c s_0$ by $\nu_g$ we have:

$$\nu_0 \nu_g R_c^2 = c s_0 \nu_g$$

since $\nu_g = c / R_c$ :

$$s_0 \nu_g = \nu_0 R_c$$

and if we use again: $\nu_0 = 1/T_0$ and: $R_c = c\, t_c$ we have:

$$\nu_g\, s_0 = c\, \frac{t_c}{T_0} = c\, \frac{\tau_0}{t_c}$$

Again, if we identify $t_c$ with $t_{UNI}$, the right hand side of this last equation is a velocity proportional to the ratio $\tau_0 / t_{UNI}$ with the proportionality constant equal to $c$. Being this velocity univocally associated to $s_0$ we may write this velocity as $v_0$ and obtain as a consequence what we may call at this point the prototypical Hubble law:

$$v_0 = H_0\, s_0$$

where $v_0$ is the galactic recession velocity as evaluated by an observer located inside our physical universe. Let us point out that the above considerations imply the following:

$$v_0 = c \frac{\tau_0}{t_{UNI}}$$

taking into account that $c/t_{UNI}$ has the dimensions of an acceleration, we may write:

$$v_0 = a_c \tau_0$$

where $a_c$ can be identified as our space-time continuum intrinsic acceleration.

**Gravitation and position**

$$\frac{4\pi^2 g_0^2}{G^2} = \frac{4\pi^2 s_0^2}{R_c^4} \quad \rightarrow \quad \frac{g_0^2}{G^2} = \frac{s_0^2}{R_c^4} \quad \rightarrow \quad R_c^2 g_0 = s_0 G$$

using again: $g_0 = \dfrac{c^2}{M_0}$ we have: $\dfrac{c^2 R_c^2}{M_0} = s_0 G \quad \rightarrow \quad M_0 = \dfrac{c^2 R_c^2}{s_0 G}$

which express the correspondence between the mass of an eigen-universe and its quantum space-time interval.

To gain further insight we can express $s_0$ and $R_c$ as follows:

$$s_0 = c \tau_0 \qquad R_c = c t_c \qquad \text{to obtain:} \qquad M_0 = \frac{c^3 t_c^2}{G \tau_0}$$

which is the relationship between the mass of the eigen-universe and its quantum proper time.

Now if we pose: $\tau_0 = t_{UNI}$ which means obviously: $s_0 = R_{UNI}$, and identify $t_c$ with $t_{UNI}$ we have:

$$M_0 = \frac{c^3 t_{UNI}}{G} = M_{UNI}$$

which shows the correlation existing between gravitation and total mass of the universe.

Another very interesting result is obtained using the relationship deduced in the momentum-position comparison: $s_0 R_0 = R_c^2$. Substituting this equation into:

$$M_0 = \frac{c^2 R_c^2}{s_0 G} \qquad \text{one obtains:} \qquad R_0 = \frac{GM_0}{c^2}$$

which is easily recognised to be our usual definition of gravitational radius $R_G$. On the other hand we know that:

$$R_0 = \frac{h}{m_0 c} \qquad \text{so:} \qquad \frac{GM_0}{c^2} = \frac{h}{m_0 c} \quad \rightarrow \quad m_0 M_0 = M_p^2$$

So that we find again the hyperbolic relationship between the mass of the particle and the mass of its eigen-universe.

**Summary of equivalent quantities**

We have found a number of equivalencies that can be usefully summarised as:

$$\frac{g_0^2}{G^2} = \frac{p_0^2}{h^2} = \frac{s_0^2}{R_c^4} = \frac{v_0^2}{c^2}$$

and in the following table:

# Z THEORY: SUMMARY OF EQUIVALENCE RELATIONSHIPS

| RELATION | PHYSIC LAW | NAME | SYNTHETIC DESCRIPTION |
|---|---|---|---|
| $h \Leftrightarrow c$ | $m_0 c^2 = h\nu_0$ | De Broglie Equivalence Law | It is the famous equivalence law between the particle and wave energy formulation postulated for the first time by L.V. De Broglie. |
| $h \Leftrightarrow G$ | $m_0 M_0 = hc/G = M_p^2$ | Planck Mass | This relation allows us to calculate the value and understand the meaning of the Planck mass as the connection element between micro-macro cosmos |
| $h \Leftrightarrow R_c^2$ | $m_0 c R_c^2 = h s_0$ | Complementary law particle-universe | Equivalence between the proper mass of a particle and correspondent quantum space-time interval. It is used to interpret the Compton wavelength of a particle as the radius of its eigen-universe. |
| $c \Leftrightarrow G$ | $\nu_0 M_0 G = c^3$ | Complementary law wave-universe | Equivalence between particle proper frequency and mass of its eigen-universe. From a physical point of view it is equivalent to De Broglie law. |
| $c \Leftrightarrow R_c^2$ | $\nu_0 R_c^2 = c s_0$ | Hubble Law | Equivalence between proper frequency of a particle and the correspondent quantum space-time interval. It implies the prototype of the cosmological Hubble Law |
| $G \Leftrightarrow R_c^2$ | $c^2 R_c^2 = M_0 G s_0$ | Gravitational Radius | Equivalence between proper mass of an eigen-universe and correspondent quantum space-time interval. The definition of gravitational radius is deduced from this equivalence relationship. |

## 3. Fundamental concepts of Z theory

In this section the fundamental concepts of Z theory are outlined.

**Eigen-particle and eigen-universe**

We need first to clarify and generalise the concepts of eigen-particle and eigen-universe, to provide a complete new picture of our physical world. This can be done by starting again from the equivalence between gravitation and position which generates the identity between gravitational radius and radius of the eigen-universe:

$$R_0 = \frac{GM_0}{c^2} \qquad R_0 = \frac{h}{m_0 c}$$

so that:

$$\frac{GM_0}{c^2} = \frac{h}{m_0 c}$$

and:

$$\frac{Gm_0}{c^2} = \frac{h}{M_0 c}$$

As a consequence we can identify the left hand side as the gravitational radius of the particle:

$$r_G = \frac{Gm_0}{c^2}$$

Interpreting the quantity $h/M_0 c$ as the radius of the eigen-particle of the universe with mass $M_0$, we have:

$$r_0 = \frac{h}{M_0 c}$$

We can summarise the concepts of eigen-particle and eigen-universe as follows:

$$R_0 = \frac{h}{m_0 c} \quad \text{eigen-universe radius of a particle with proper mass } m_0$$

$$m_0 = \frac{h}{R_0 c} \quad \text{eigen-particle mass of a universe with radius } R_0$$

$$r_0 = \frac{h}{M_0 c} \quad \text{eigen-particle radius of a universe with proper mass } M_0$$

$$M_0 = \frac{h}{r_0 c} \quad \text{eigen-universe mass of a particle with radius } r_0$$

$$m_0 M_0 = \frac{hc}{G} = M_p^2 \quad \text{hyperbolic law particle/eigen-universe mass}$$

We should note that in the above relationships the quantities related to eigen-particles or eigen-universes are completely interchangeable.
This introduces the concept of **UP symmetry** (Universe-Particle symmetry) as a natural (logical) consequence.
In other words, UP symmetry assures us that we can describe physical reality in terms of either particles or universes, in an equivalent manner.

**The eigen-particle of the universe**

For example, we may now calculate the mass of the eigen-particle of our universe. This is done using the second expression of the above list. With the following values of the involved quantities:

$h = 6.6256 \; 10^{-34}$ J s $\quad c = 2.998 \; 10^8$ m/s $\quad R_0 \cong 1.42 \; 10^{26}$ m

the result is: $\quad m_0 = 1.556 \; 10^{-68}$ Kg.

We can now also calculate easily the equivalent mass of the universe using the last of the above laws:

$$M_{UNI} = M_p^2 / m_0 \cong 1.9126 \; 10^{53} \text{ Kg}$$

and observe that this result is in line with the predictions of the cosmological models derived from general relativity theory.

The particle with the above proper mass $m_0$ can be naturally identified as the quantum projection of our universe and in this sense we may recognise it as the quantum of gravity, i.e. the graviton ($m_0 = m_g$).

In other words:

> **the graviton is the smallest massive particle existing in our universe, its Compton wavelength is equal to the radius of the universe.**

**The role of the Planck quantities**

One of the relationships already found, relates the mass of the particle to the mass of its eigen-universe:

$$m_0 M_0 = \frac{hc}{G} = M_p^2$$

On the other hand we can try the same approach with the radius of the particle and the radius of its eigen-universe, which is:

$$r_0 R_0 = \frac{h}{M_0 c} \frac{h}{m_0 c} = \frac{h^2}{M_p^2 c^2} = \frac{Gh}{c^3} = L_p^2$$

Dividing the above by $c^2$, we obtain: $t_r T_R = T_p^2$.

If $m_0$ is the mass of the eigen-particle of the universe: $m_0 = M_{MIC}$, where $M_{MIC}$ is the proper mass of the microcosmos.

It follows that: $M_0 = M_{MAC}$, where $M_{MAC}$ is the proper mass of the macrocosmos (the mass of the universe). Using similar considerations for lenghts ($L_{MIC}, L_{MAC}$) and times ($T_{MIC}, T_{MAC}$) we have:

$$M_{MIC} \cdot M_{MAC} = M_p^2$$
$$L_{MIC} \cdot L_{MAC} = L_p^2$$
$$T_{MIC} \cdot T_{MAC} = T_p^2$$

**The Planck quantities are the point of equilibrium between microcosmos and macrocosmos and show the existence of this hyperbolic symmetry in the universe.**

More insight can be gained by expressing the above as follows and calculating the values in the case of our universe:

$$\frac{M_0}{M_p} = \frac{M_p}{m_0} \cong 3.5 \; 10^{60} \qquad \frac{R_0}{L_p} = \frac{L_p}{r_0} \cong 3.5 \; 10^{60} \qquad \frac{T_R}{T_p} = \frac{T_p}{t_r} \cong 3.5 \; 10^{60}$$

or: 
$$\frac{M_0}{m_0} = \frac{R_0}{r_0} = \frac{T_R}{t_r} \cong 1.229 \; 10^{121}$$

where the first constant can be considered representative of the "distance" between our universe and the Planck universe, while the second one is representative of the "distance" between our universe and its eigen-particle, the graviton.

It has to be noted that the quantities $t_r$ and $T_R$ have to be properly interpreted as the time interval needed by the speed of light to cover a distance equal to the radius of the particle $r_0$ and the radius of the eigen-universe $R_0$. In the case of the graviton, $t_r$ and $T_R$ are continuously varying, $t_r$ is decreasing and $T_R$ is increasing. In the case of a "normal" particle, and supposing this mass is constant, such quantities are obviously constants.
On the other hand, if we consider the already deduced relationship:
$s_0 R_0 = R_c^2$ and divide it by $c^2$, we obtain: $\tau_0 T_0 = t_c^2$.
This relationship defines the quantum proper time of the particle $\tau_0$.
$T_0$ instead is constant and equal to $T_R$. This for a normal particle. For the graviton $s_0 = R_0$, so that the above expression reduces to: $T_0 \, T_0 = t_c^2 = T_0^2$.

**Explicit expressions of the four-operators and four-vectors**

We start with the already known four-operator and four-vector.

**Momentum**

In this case $K_{QO} = h$, substituting:

$$[\mathbf{p}] = \frac{h}{2\pi i} \frac{\partial}{\partial s_n}\bigg|_{n=1,4} \qquad \mathbf{p} \rightarrow [\mathbf{p}] = \frac{h}{2\pi i}(\frac{\partial}{c\partial t}, \frac{\partial}{i\partial x}, \frac{\partial}{i\partial y}, \frac{\partial}{i\partial z})$$

from which we are able to deduce the Schrödinger operators in the following form:

$$\hat{E} = \frac{h}{2\pi i}\frac{\partial}{\partial t} \qquad \hat{p} = -\frac{h}{2\pi i}\frac{\partial}{\partial x_\alpha}\bigg|_{\alpha=1,3}$$

This is the common definition of the energy-momentum operator, except that the sign is reversed. The associated four-vector is:

$$[\mathbf{p}]\Psi = \frac{h}{hc}E\,(1, i\beta_x, i\beta_y, i\beta_z)\,\Psi = \frac{E}{c}(1, i\beta_x, i\beta_y, i\beta_z)\,\Psi$$

$$[\mathbf{p}]\Psi = (mc, imv_{px}, imv_{py}, imv_{pz})\,\Psi$$

which shows that the momentum four-vector is an eigen-vector of the four-operator momentum. The expression of the four-vector is:

$$\mathbf{p} = \frac{E}{c}(1, i\beta_x, i\beta_y, i\beta_z)$$

and the associated algebraic equation (the relativistic energy equation) obtained with the dot product of the momentum four-vector with itself is as a consequence:

$$E^2/c^2 - p^2 = m_0^2 c^2$$

**Frequency**

In this case $K_{QO} = c$, substituting:

$$[\nu] = \frac{c}{2\pi i} \frac{\partial}{\partial s_n}\bigg|_{n=1,4} \qquad \nu \rightarrow [\nu] = \frac{c}{2\pi i}(\frac{\partial}{c\partial t}, \frac{\partial}{i\partial x}, \frac{\partial}{i\partial y}, \frac{\partial}{i\partial z})$$

The associated four-vector is:

$$[\nu]\Psi = \frac{c}{hc}E(1, i\beta_x, i\beta_y, i\beta_z)\Psi = \frac{E}{h}(1, i\beta_x, i\beta_y, i\beta_z)\Psi$$

so that the final expression for the frequency four-vector is:

$$\nu = \frac{E}{h}(1, i\beta_x, i\beta_y, i\beta_z)$$

From this expression it is clear why we identified this quantity as a frequency: in fact $E/h$ has the dimensions of a frequency. As a consequence the intrinsic frequency four-vector is:

$$\nu = \nu(1, i\beta_x, i\beta_y, i\beta_z)$$

We may also associate to each particle an intrinsic acceleration by multiplying the intrinsic frequency four-vector by $c$, as follows:

$$\boldsymbol{a} = \nu c(1, i\beta_x, i\beta_y, i\beta_z)$$

and note that the ratio between the momentum and acceleration four-vectors is equal to a constant: $\boldsymbol{p}/\boldsymbol{a} = h/c^2$ as well as : $\boldsymbol{p}/\boldsymbol{\nu} = h/c$

Finally, the frequency algebraic equation is:

$$\nu^2(1 - \beta^2) = \nu_0^2$$

## Gravitation

In the case of $K_{QO} = G$:

$$[\mathbf{g}] = \frac{G}{2\pi i} \frac{\partial}{\partial s_n}\bigg|_{n=1,4} \qquad \mathbf{g} \to [\mathbf{g}] = \frac{G}{2\pi i}\left(\frac{\partial}{c\partial t}, \frac{\partial}{i\partial x}, \frac{\partial}{i\partial y}, \frac{\partial}{i\partial z}\right)$$

The associated four-vector is:

$$[\mathbf{g}]\Psi = \frac{G}{hc} E\, (1, i\beta_x, i\beta_y, i\beta_z)\Psi = \frac{E}{\frac{hc}{G}}(1, i\beta_x, i\beta_y, i\beta_z)\Psi$$

so that we may express the gravitation four-vector using the Planck mass as:

$$\mathbf{g} = \frac{E}{M_p^2}(1, i\beta_x, i\beta_y, i\beta_z)$$

<u>The Planck mass participates directly to the structure of the gravitation four-vector</u>.

If one defines: $g = \frac{E}{M_p^2}$, the algebraic equation is: $g_0^2 = g^2(1 - \beta^2)$.

We may also consider the following relationships:

$$g = \frac{E}{M_p^2} = \frac{mc^2}{M_p^2} = \frac{c^2}{M}$$

where $M$ is a mass associated to the particle of mass $m$ such that $mM = M_p^2$. This above relationship is obviously true in the reference system of the particle so: $m_0 M_0 = M_p^2$. As a consequence we found again the mass $M_0$ as the proper mass of the eigen-universe associated to the particle of proper mass $m_0$.

**Position**

The fourth four-operator is obtained using the evolutionary variable $R_c$, as follows:

$$[s] = \frac{R_c^2}{2\pi i} \frac{\partial}{\partial s_n}\bigg|_{n=1,4} \quad s \to [s] = \frac{R_c^2}{2\pi i} \left( \frac{\partial}{c\partial t}, \frac{\partial}{i\partial x}, \frac{\partial}{i\partial y}, \frac{\partial}{i\partial z} \right)$$

The associated four-vector is:
$$s = \frac{R_c^2}{hc} E \, (1, i\beta_x, i\beta_y, i\beta_z)$$

To get an explicit form of the four-vector we rewrite it using $G$:

$$s = \frac{E}{G\dfrac{hc}{G}\dfrac{1}{R_c^2}} (1, i\beta_x, i\beta_y, i\beta_z) \quad \text{so:} \quad s = \frac{E}{G\dfrac{M_p M_p}{R_c^2}} (1, i\beta_x, i\beta_y, i\beta_z)$$

and we can now recognise in the quantity: $\quad G\dfrac{M_p M_p}{R_c^2} = G\dfrac{m_0 M_0}{R_c^2}$

(which has the dimensions of a force) what we may consider the prototype of the universal gravitational law.
As a consequence we can write:

$$s = \frac{E}{F_{GC}} (1, i\beta_x, i\beta_y, i\beta_z)$$

A very important result may be deduced using the dot product. We already know that:

$$s \circ s = s_0^2 \quad \text{so:} \quad s_0^2 = \frac{E_0^2}{F_{GC}^2}$$

So we may write: $\quad F_{GC}\, s_0 = E_0$

where the prototype of our usual definition of work is recognised.

Additionally using: $E_0 = m_0 c^2$ and $s_0 = c\tau_0$ we obtain:

$$\frac{c^2 \tau_0^2}{m_0^2 c^4} = \frac{1}{F_{GC}^2}$$

But the quantity $c/\tau_0$ is an intrinsic squared acceleration. If we define:

$$a_0^2 = \frac{c^2}{\tau_0^2} \quad \text{we have: } m_0^2 a_0^2 = F_{GC}^2$$

which may be interpreted as the prototype of the equivalence principle.

Finally, it may be shown that the space-time continuum equation is the algebraic equation associated to the dot product of the position four-vector with itself.

In fact, using the relationship $m_0 = \dfrac{h}{c R_c^2} s_0$ in the expression of the dot product of the position four-vector with itself, we obtain:

$$c^2 t^2 - x^2 - y^2 - z^2 = s_0^2$$

and we can then write:

$$\mathbf{s} = (ct, ix, iy, iz)$$

As a consequence, the associated algebraic equation of the position operator is the quantised Einstein-Minkowski space-time continuum, i.e. the quantum space-time interval:

$$\mathbf{s} \circ \mathbf{s} = c^2 t^2 - x^2 - y^2 - z^2 = s_0^2$$

## Gravity and Electromagnetism

Before summarising the four-operators and four-vectors, we now develop some considerations on the basis of the form of the four-vector position just presented.

We begin again from the logical functional prototype of what we will call from now on the Gravitational (space-time) Continuum Force ($F_{GC}$), and write down its value for the electron. In this way we evaluate the gravitational force between the electron and its eigen-universe:

$$F_{GC} = G \frac{m_e M_e}{R_c^2} = G \frac{M_p^2}{R_c^2} \cong 9.8515 \; 10^{-78} \; N$$

We may now ask ourselves the following question:
"What is the electrostatic force between two electrons at a distance equal to the radius of our space-time continuum?" The answer is given by the Coulomb law as:

$$F_{ELS} = K_c \frac{e\,e}{R_c^2} = 1.144 \; 10^{-80} \; N$$

Where $K_c$ is the Coulomb constant and $e$ is the electronic charge.

If we now introduce the fine structure constant defined as:

$$\alpha = \frac{K_c e^2}{\hbar c}$$

it is easy to show that the above two forces are related by the following relationship:

$$F_{ELS} = F_{GC} \frac{\alpha}{2\pi}$$

So that the absolute value of the electrostatic force is simply the gravitational continuum force multiplied by the fine structure constant divided by $2\pi$.

We may introduce at this point the Z interpretation of this result:

**The electrostatic force is the gravitational continuum force between the first electron and the eigen-universe of the second electron. The coupling constant is the fine structure constant.**
If the above interpretation is correct, the electric charge is the direct consequence of the particle-universe symmetry.

We would like to stress that without previous knowledge of the electrostatic force, but only on the basis of Z theory, one could however predict the existence of an additional force besides the gravitational one.

**Summary of four-operators and four-vectors**

We now present a summary of four-operators and four-vectors as well as their associated algebraic equations:

$$[\mathbf{p}] \, \Psi = \frac{h}{2\pi i} \frac{\partial}{\partial s_n} \Psi = \mathbf{p} \, \Psi = \frac{E}{c} (1, i\beta_x, i\beta_y, i\beta_z) \Psi$$

$$[\mathbf{\nu}] \, \Psi = \frac{c}{2\pi i} \frac{\partial}{\partial s_n} \Psi = \mathbf{\nu} \, \Psi = \frac{E}{h} (1, i\beta_x, i\beta_y, i\beta_z) \Psi$$

$$[\mathbf{g}] \, \Psi = \frac{G}{2\pi i} \frac{\partial}{\partial s_n} \Psi = \mathbf{g} \, \Psi = \frac{E}{M_p^2} (1, i\beta_x, i\beta_y, i\beta_z) \Psi$$

$$[\mathbf{s}] \, \Psi = \frac{R_c^2}{2\pi i} \frac{\partial}{\partial s_n} \Psi = \mathbf{s} \, \Psi = \frac{E}{F_{GC}} (1, i\beta_x, i\beta_y, i\beta_z) \Psi$$

so that:

$$E(1, i\beta_x, i\beta_y, i\beta_z) = c\mathbf{p} = h\mathbf{\nu} = M_p^2 \mathbf{g} = F_{Gc} \mathbf{s} = \mathbf{E}$$

We note the presence of four basic quantities in the previous equations: two of them already included in the set of axioms while the remaining have been deduced.

We note that **the set of axioms of Z theory implies the existence of the Planck mass and of the gravitational space-time continuum force prototype**.

The above defined energy vector $\boldsymbol{E} = E\,(1, i\beta_x, i\beta_y, i\beta_z)$ is the physical aspect which is able to unify the different descriptions of the physic world and in particular: the dynamical (motion), vibrational (frequency), gravitational (mass) and space-time continuum (position) aspects.

We should finally summarise the four quadratic algebraic equations associated to the dot product of each four-vector with itself:

$h\ \ :\ E^2 - p^2 c^2 = E_0^2$ relativistic energy equation

$c\ \ :\ \nu^2 (1 - \beta^2) = \nu_0^2$ relativistic frequency

$G\ \ :\ M^2 = M_0^2\,(1 - \beta^2)$ relativistic mass of the eigen-universe

$R_c^2\ :\ c^2 t^2 - x^2 - y^2 - z^2 = s_0^2$ quantum space-time interval

## 4. Quantum-relativistic algebra

We would now like to explore some of the combinatorial possibilities offered by the four-vectors of Z theory. In the following, we are going to present some of results of the analysis.
A more detailed investigation has to be made using the four-operators, however this is not part of the present work and will be the object of a separate investigation.
The dot products of four-vectors with themselves generate the following relativistic invariants:

$$\boldsymbol{p} \circ \boldsymbol{p} = \frac{E_0^2}{c^2} = m_0^2 c^2 = p_0^2 \qquad \boldsymbol{v} \circ \boldsymbol{v} = \frac{E_0^2}{h^2} = v_0^2$$

$$\boldsymbol{g} \circ \boldsymbol{g} = \frac{E_0^2}{M_p^4} = \frac{c^4}{M_0^2} = g_0^2 \qquad \boldsymbol{s} \circ \boldsymbol{s} = \frac{E_0^2}{F_{GC}^2} = s_0^2$$

But we can also calculate other relativistic invariants using "mixed" dot products as follows:

$$\boldsymbol{p} \circ \boldsymbol{v} = \frac{E}{c}(1, i\vec{\beta}) \frac{E}{h}(1, i\vec{\beta}) = \frac{E^2}{ch}(1-\beta^2) = \frac{E_0^2}{ch} = m_0^2 \frac{c^3}{h}$$

$$\boldsymbol{p} \circ \boldsymbol{g} = \frac{E}{c}(1, i\vec{\beta}) \frac{E}{M_p^2}(1, i\vec{\beta}) = E_0^2 / \frac{c^2 h}{G} = m_0^2 \frac{c^4 G}{c^2 h} = m_0^2 \frac{c^2 G}{h}$$

$$\boldsymbol{p} \circ \boldsymbol{s} = \frac{E}{c}(1, i\vec{\beta}) \frac{E}{F_{GC}}(1, i\vec{\beta}) = E_0^2 / c \frac{GM_p^2}{R_c^2} = \frac{E_0^2 R_c^2}{c^2 h} = m_0^2 \frac{c^2 R_c^2}{h}$$

$$\boldsymbol{v} \circ \boldsymbol{g} = \frac{E}{h}(1, i\vec{\beta}) \frac{E}{M_p^2}(1, i\vec{\beta}) = E_0^2 / \frac{h^2 c}{G} = m_0^2 \frac{c^3 G}{h^2}$$

$$\boldsymbol{v} \circ \boldsymbol{s} = \frac{E}{h}(1, i\vec{\beta}) \frac{E}{F_{GC}}(1, i\vec{\beta}) = E_0^2 / h \frac{GM_p^2}{R_c^2} = \frac{E_0^2 R_c^2}{c h^2} = m_0^2 \frac{c^3 R_c^2}{h^2}$$

$$\boldsymbol{g} \circ \boldsymbol{s} = \frac{E}{M_p^2}(1, i\vec{\beta}) \frac{E}{F_{GC}}(1, i\vec{\beta}) = E_0^2 / M_p^2 \frac{GM_p^2}{R_c^2} = m_0^2 \frac{c^2 G R_c^2}{h^2}$$

Please note that the final results are always expressed by the product of a proper mass with some of the fundamental constants or quantity.

It is very interesting to obtain the above dot products for two particular cases. If the particle mass is the graviton mass:

$$\boldsymbol{p} \circ \boldsymbol{s} = h \qquad \boldsymbol{v} \circ \boldsymbol{s} = c \qquad \boldsymbol{g} \circ \boldsymbol{s} = G$$

$$\boldsymbol{p} \circ \boldsymbol{v} = \frac{hc}{R_c^2} \qquad \boldsymbol{p} \circ \boldsymbol{g} = \frac{hG}{R_c^2} \qquad \boldsymbol{v} \circ \boldsymbol{g} = \frac{cG}{R_c^2}$$

while if the particle mass is equal to the Planck mass we obtain:

$$\boldsymbol{p} \circ \boldsymbol{s} = \frac{h R_c^2}{L_p^2} \qquad \boldsymbol{v} \circ \boldsymbol{s} = \frac{c R_c^2}{L_p^2} \qquad \boldsymbol{g} \circ \boldsymbol{s} = \frac{G R_c^2}{L_p^2}$$

$$\boldsymbol{p} \circ \boldsymbol{v} = \frac{c^4}{G} \qquad \boldsymbol{p} \circ \boldsymbol{g} = \frac{c^4}{c} \qquad \boldsymbol{v} \circ \boldsymbol{g} = \frac{c^4}{h}$$

The above expressions point out the symmetries existing in the two different situations.

Additional four-vectors can be generated by multiplication of constants with the four fundamental four-vectors. This is the case, for example, of the mass and acceleration four-vector:

$$\boldsymbol{m} = \frac{\boldsymbol{p}}{c} = \frac{mc(1, i\vec{\beta})}{c} = m(1, i\vec{\beta}) = m(1, i\beta_x, i\beta_y, i\beta_z)$$

$$\boldsymbol{a} = c\boldsymbol{v} = c\nu(1, i\vec{\beta}) = c\nu(1, i\beta_x, i\beta_y, i\beta_z)$$

or of the very important energy four-vector, which may take the form:

$$\boldsymbol{E} = c\boldsymbol{p} = mc^2(1, i\vec{\beta}) = mc^2(1, i\beta_x, i\beta_y, i\beta_z)$$

or: $$\boldsymbol{E} = h\boldsymbol{\nu} = h\nu(1, i\vec{\beta}) = h\nu(1, i\beta_x, i\beta_y, i\beta_z)$$

At this point, we may leave to the reader the exploration of other interesting possibilities.

Some additional considerations are now developed only for the case of momentum and frequency dot product. Dimensional analysis tells us that this dot product is a force:

$$F = \mathbf{p} \circ \mathbf{v} = \frac{E}{c^2}(1, i\vec{\beta}) \frac{Ec}{h}(1, i\vec{\beta}) = m(1, i\vec{\beta}) \, a(1, i\vec{\beta}) = \mathbf{m} \circ \mathbf{a}$$

which we may consider as the logical functional prototype of Newton's second law. But we may continue our analysis taking into account that:

$$\mathbf{p} \circ \mathbf{v} = m_0^2 \frac{c^3}{h} \quad \text{so that:} \quad F_0 = m_0 \, a_0 = m_0^2 \frac{c^3}{h} \quad \rightarrow \quad m_0 = a_0 \frac{h}{c^3}$$

and obtain:

$$\boxed{F_0 = a_0^2 \frac{h}{c^3}}$$

this shows a quadratic relationship between the proper force and the proper acceleration.

If we use $a_0 = a_c = c/t_c$, we have:

$$F_0 = a_c^2 \frac{h}{c^3} = \frac{c^4}{R_c^2} \frac{h}{c^3} = \frac{ch}{R_c^2} = \frac{GM_p^2}{R_c^2} = F_{GC}$$

so that:
$$F_0 = m_g \, a_c = F_{Ig} = F_{GC}$$

In other words the proper force $F_0$ is equal to the inertial intrinsic force of the graviton and to the gravitational continuum force.

As a consequence, the Einstein equivalence principle may be expressed as:

$$\boxed{F_{Ig} = F_{GC}}$$

A similar relationship exists for the energy:

$$\boxed{E_g = E_{GC}}$$

showing that the internal (or intrinsic) energy of the graviton is equal to the gravitational continuum energy, this because:

$$E_g = m_g c^2 \qquad E_{GC} = \frac{G M_p^2}{R_c} \quad \rightarrow \quad \frac{h}{c R_c} c^2 = G \frac{hc}{G} \frac{1}{R_c}$$

In general:

$$E_0 = m_0 c^2 = \frac{h}{c R_0} c^2 = \frac{hc}{R_0} \qquad E_{G0} = \frac{G m_0 M_0}{R_0} = \frac{G M_p^2}{R_0} = \frac{hc}{R_0}$$

so that the internal energy of a particle of proper mass $m_0$ is formally equal to the gravitational energy of its eigen-universe.
We will call this statement **Z-equivalence principle**.
<u>Following the Z-equivalence principle, the inertial properties of a particle completely corresponds to the gravitational properties of its eigen-universe</u>.

We like to conclude this part of the paper showing the use of some of the dot products already calculated. We already know that:

$$\Psi = e^{\frac{2\pi i}{h}(\mathbf{p} \circ \mathbf{s})}$$

But, to complete the picture, we should also propose the following forms:

$$\Psi = e^{\frac{2\pi i}{c}(\mathbf{v} \circ \mathbf{s})} \qquad \Psi = e^{\frac{2\pi i}{c^2}(\mathbf{a} \circ \mathbf{s})} \qquad \Psi = e^{\frac{2\pi i}{G}(\mathbf{g} \circ \mathbf{s})}$$

so that now all dot products are explicitely shown in the basic structure of the wave function.

## 5. On the uncertainty principle

Let's now look at where the Z theory stands in relation to the uncertainty principle.

It is worthwhile first pointing out that, in the framework of quantum mechanics, the Heisenberg uncertainty principle (as well as other uncertainty relationships) can be directly derived from the non-commutative algebra of the operators associated to physicals observables. In this sense, the Heisenberg uncertainty principle is a consequence of the form of the operators as defined by standard quantum mechanics. The operator associated to momentum is a differential operator, while the position operator takes the form of the position variable. Mathematically, this produces the impossibility for such operators to be commutative.

In marked contrast, **the quantum-relativistic operators defined by Z theory clearly obey a completely commutative algebra**, because position and momentum are represented by differential operators in this theory, so that the commutator obviously equals zero.

To clarify the above statements, we form the commutator of the four-operators in the case of quantum mechanics and in Z theory.

The quantum mechanics position and momentum four-operators can be defined, following the notations used in this work, as follows:

$$[r] = (ct, ix, iy, iz) \qquad [p] = \frac{h}{2\pi i}(\frac{\partial}{c\partial t}, \frac{\partial}{i\partial x}, \frac{\partial}{i\partial y}, \frac{\partial}{i\partial z})$$

if we write the commutator of the four-operators we have:

$$\{[p], [r]\}\Psi = \frac{h}{2\pi i}((\frac{\partial}{c\partial t}, \frac{\partial}{i\partial x}, \frac{\partial}{i\partial y}, \frac{\partial}{i\partial z})(ct, ix, iy, iz))\Psi +$$

$$- (ct, ix, iy, iz)\frac{h}{2\pi i}(\frac{\partial}{c\partial t}, \frac{\partial}{i\partial x}, \frac{\partial}{i\partial y}, \frac{\partial}{i\partial z})\Psi$$

Note that in the above expression the momentum operator has been applied to the position operator and viceversa, so that the result of the commutator will be a four dimensional quantity.

We now calculate the commutator:

$$\{[\mathbf{p}], [\mathbf{r}]\} \Psi = \frac{h}{2\pi i} \left( \frac{c\partial t}{c\partial t}, \frac{i\partial x}{i\partial x}, \frac{i\partial y}{i\partial y}, \frac{i\partial z}{i\partial z} \right) \Psi +$$

$$+ (ct, ix, iy, iz) \frac{h}{2\pi i} \left( \frac{\partial \Psi}{c\partial t}, \frac{\partial \Psi}{i\partial x}, \frac{\partial \Psi}{i\partial y}, \frac{\partial \Psi}{i\partial z} \right) +$$

$$- (ct, ix, iy, iz) \frac{h}{2\pi i} \left( \frac{\partial \Psi}{c\partial t}, \frac{\partial \Psi}{i\partial x}, \frac{\partial \Psi}{i\partial y}, \frac{\partial \Psi}{i\partial z} \right)$$

to obtain:

$$\boxed{\{[\mathbf{p}], [\mathbf{r}]\} \Psi = \frac{h}{2\pi i} (1,1,1,1) \Psi}$$

which is the expression of the four-dimensional commutator deduced from the definition of the operators of standard quantum mechanics.

We now perform the same process using the quantum-relativistic operators of Z theory, defined as follows:

$$[\mathbf{s}] = \frac{R_c^2}{2\pi i} \left( \frac{\partial}{c\partial t}, \frac{\partial}{i\partial x}, \frac{\partial}{i\partial y}, \frac{\partial}{i\partial z} \right)$$

$$[\mathbf{p}] = \frac{h}{2\pi i} \left( \frac{\partial}{c\partial t}, \frac{\partial}{i\partial x}, \frac{\partial}{i\partial y}, \frac{\partial}{i\partial z} \right)$$

and obtain in this case:

$$\boxed{\{[\mathbf{p}], [\mathbf{s}]\} \Psi = \frac{h R_c^2}{4\pi^2} (0,0,0,0) \Psi = (0,0,0,0)}$$

This last expression shows that the quantum-relativistic operators commute, so we have been able to show that:

**the "Z reality level" does not embody any quantum uncertainty**

This fact may give rise to the following question: what happens, in the framework of Z theory, to the quantum mechanics uncertainty principle?

To answer this basic question we have to analyse the relationship between the eigen-particle and eigen-universe already introduced in Z theory in more detail.

This analysis will bring us to a new interpretation and different understanding of the quantum mechanics uncertainty.

Let us first write the Heisenberg uncertainty relationship as:

$$\Delta r \, \Delta p = h$$

and move now to the reference system of the particle [7] to express it as:

$$\Delta r \, m_0 c = h$$

which defines the minimum possible uncertainty on the particle position.

Now, we can compare this last expression with the relationship between the mass of a particle and the radius of its eigen-universe:

$$m_0 c R_0 = h$$

and realise that the system of such two equations gives rise to the identity:

$$\Delta r = R_0$$

The uncertainty on the particle position is equal to the radius of its eigen-universe.

In other words the quantum uncertainty is a consequence of the existence of the complementary law between eigen-particle and eigen-universe, witness of the impossibility to separate the concept of elementary particle and eigen-universe.

The quantum uncertainty of the position of a massive particle can not be reduced below the radius of its eigen-universe. In the same way the quantum uncertainty on momentum can not be reduced below the value $m_0 c$ i.e. below the "proper" momentum of the particle.

To further explain the position of Z theory on quantum uncertainty we can say that: if we try to determine the particle position with a precision below the extension of its eigen-universe we are forcing a change of the particle mass so that we are modifying and effectively observing another particle with respect to what we initially want to observe.

The importance of the eigen-particle, eigen-universe duality is now completely clear and highlights a new **physical** interpretation of the quantum uncertainty.

Let us summarise our position in the following statement:

**The quantum uncertainty is a consequence of the forced confinement of a particle with respect to the radius of its eigen-universe: the response of the physical reality to the conceptual separation of the "particle" aspect with respect to the "universe" aspect.**

# 6. The meaning of $|\Psi|^2$

In this last part of the work, we would like to face one of the basic problems of quantum mechanics, the meaning of the wave function. In fact quantum mechanics provides an interpretation for the squared module of the wave function (as the probability to find the particle in a specific space portion), while no specific physical interpretation is provided for the wave function itself.
In the following we will try to suggest some developments that will bring us to understand the deep physical meaning of the squared module of a quantity which is apparently without physical meaning.

We are in a position to answer this fundamental question for the case of a free particle. We have to go back to the formulation of differential equations associated to the momentum and position operators of Z theory:

$$\frac{\partial^2 \Psi}{\partial x^2} + \frac{\partial^2 \Psi}{\partial y^2} + \frac{\partial^2 \Psi}{\partial z^2} - \frac{1}{c^2}\frac{\partial^2 \Psi}{\partial t^2} = \frac{4\pi^2 m_0^2 c^2}{h^2} \Psi$$

$$\frac{\partial^2 \Psi}{\partial x^2} + \frac{\partial^2 \Psi}{\partial y^2} + \frac{\partial^2 \Psi}{\partial z^2} - \frac{1}{c^2}\frac{\partial^2 \Psi}{\partial t^2} = \frac{4\pi^2 s_0^2}{R_c^4} \Psi$$

We can transform now the above equations using the d'Alembert operator as follows:

$$\Box^2 \Psi = \frac{4\pi^2 m_0^2 c^2}{h^2} \Psi \qquad \Box^2 \Psi = \frac{4\pi^2 s_0^2}{R_c^4} \Psi$$

if we now restrict our analysis to the wave function of the free particle, the following identity holds:

$$\Psi^* = \Psi^{-1}$$

so that we can rearrange the two equations as:

$$\Psi/\Box^2\Psi = \frac{h^2}{4\pi^2 m_0^2 c^2} \qquad \Psi^*\Box^2\Psi = \frac{4\pi^2 s_0^2}{R_c^4}$$

we can now multiply the two equations together to obtain:

$$\Psi\Psi^* = \frac{h^2 s_0^2}{m_0^2 c^2 R_c^4}$$

and using the relationship between the quantum space-time interval and the eigen-universe radius $s_0^2 R_0^2 = R_c^4$ we finally have:

$$\Psi\Psi^* = \frac{h^2}{m_0^2 c^2 R_0^2}$$

This last equation shows us that only the squared module of the wave function gives us access to the totality of information on the physical entity: in fact it summarises both the particle massive aspects ($m_0$) as well as its space-time continuum aspects ($R_0$).

It has to be noted that the starting differential equations represent the propagation of a physical entity with mass (momentum operator) and the propagation of its own eigen-universe (position operator). Both equations have to be considered simultaneously to obtain what we call "physical reality".

Again, this brings us to the conclusion that a particle cannot be separated from its own space-time continuum, and only the squared module of the wave function is capable of summarising both aspects of the physical reality.

It is now clear that the wave function is only part of the information on the physical entity. Such information has to be completed with that contained in $\Psi^*$ to have an exhaustive knowledge of our physical reality.

## Acknowledgements

Many thanks to Daniela Migliorini and Sandra Hitken for the correction of the English manuscript as well as to Enrico Silvestri for the useful suggestions on the draft version of the paper.

## Bibliography

1 - Pietro Giorgio Zerbo - Teoria Z – Verso una sintesi del mondo fisico
    2003 – Le Mani – Microart's Edizioni, Genova – ISBN 88-8012-266-5

2 - Richard P. Feynman - "The Feynman Lectures of Physics"
    Addison-Wesley Longman, 1989

3 - Herbert Goldstein - "Classical Mechanics"
    Addison-Wesley Publishing Company,1965

4 - Alberto Bandini Buti - "Meccanica Ondulatoria e Quantistica"
    Editoriale Delfino, Milano, 1962

5 - P.A.M. Dirac "The Principles of Quantum Mechanics"
    Oxford University Press, $4^{th}$ edition, 1958

6 - A. S. Davydov - "Kvantovaja Mechanika" - MIR, Moscow, 1981

7 - L.D. Landau, E. M. Lifšits - "Reljativistskaja Kvantovaj Teorija"
    MIR, Moscow, 1978

8 - L.D. Landau, E. M. Lifšits -"ТЕОРИЯ ПОЛЯ", MIR, Moscow, 1966

9 - Roland Omnès - "The Interpretation of Quantum Mechanics"
    Princeton University Press, Princeton, New Jersey, 1994

10 - J.Butterfield, C.Isham, S.Weinstein, C.Rovelli, E.Witten, R.Weingard, W.G.Unruh, J.C.Baez, J.B.Barbour, G.Belot, J.Earman, H.R.Brown, O.Pooley, S.Goldstein, S.Teufel, R.Penrose J.Christian
    *"Physics Meets Philosophy at the Planck Scale"*
    eds. C. Callender and N. Huggett, Cambridge University Press, 1999

11 - L. De Broglie L., E. Schrodinger , W. Heisenberg
    "Onde e particelle in armonia" - Jaca Book , Milano, 1991

12 - Roger Penrose – "The Emperor's New Mind"
    Oxford University Press, 1989

13 - David Z. Albert – "Quantum Mechanics and Experience"
    Harvard University Press, 1992

14 - Gian Carlo Ghirardi – "Un'Occhiata alle Carte di Dio"
    -Gli Interrogativi che la Scienza Moderna Pone all'Uomo-
    Il Saggiatore, Milano, 1997

15 - Luigi Accardi – "Urne e Camaleonti" - Dialogo sulla
    Realtà, le Leggi del Caso e l'Interpretazione della Teoria
    Quantistica - Il Saggiatore, Milano, 1997

16 - Steven Weinberg – "Dreams of a Final Theory"
    Pantheon Books, New York, 1993

17 - Brian R. Greene – "The Elegant Universe"
    W. W. Norton&Company, 1999

18 - Abraham Pais – "Subtle Is the Lord …"
    - The Science and The Life of Albert Einstein -
    Oxford University Press, 1982

19 - Matt Visser -"Mass for the graviton"
    Gen.Rel.Grav. 30 (1998) 1717-1728 - http://arxiv.org - gr-qc/9705051

20 - J. C. Baez - "Higher dimensional algebra and Planck-scale Physics"
    http://arxiv.org - gr-qc/9902017)

21 - I. Burud, F. Courbain, P. Magain, C. Lidman et alt.
    "An Optical Time-Delay for the Lensed BAL Quasar HE 2149 -2745"
    http://arxiv.org – astro-ph/0112225 v1, 10 December 2001

**Symbols**

| | |
|---|---|
| $a_0$ | proper acceleration |
| $a_c$ | space-time continuum acceleration |
| $c$ | speed of light |
| $e$ | electronic charge |
| $E$ | energy |
| $E_0$ | proper energy |
| $E_g$ | graviton proper energy |
| $E_G$ | gravitational energy |
| $E_{GC}$ | space-time continuum gravitational energy |
| $F$ | force |
| $F_0$ | proper force |
| $F_{ELS}$ | electrostatic force |
| $F_G$ | gravitational force |
| $F_{GC}$ | space-time continuum gravitational force |
| $F_{Ig}$ | inertial intrinsic force of the graviton |
| $g_0$ | proper gravitation |
| $G$ | gravitational constant |
| $h$ | Planck constant |
| $\hbar$ | $h/2\pi$ |
| $H_0$ | Hubble constant |
| $i$ | imaginary unit |
| $k$ | generic constant |
| $K_C$ | Coulomb constant |
| $K_{QO}$ | constant of the four-operator **[QO]** |
| $L_{MAC}$ | macrocosmos length |
| $L_{MIC}$ | microcosmos length |
| $L_p$ | Planck length |
| $m$ | relativistic mass |
| $m_0$ | proper mass |
| $m_e$ | electronic mass |
| $m_g$ | graviton mass |
| $M$ | eigen-universe mass |
| $M_0$ | eigen-universe proper mass |
| $M_c$ | space-time continuum mass |
| $M_e$ | electronic eigen-universe mass |
| $M_G$ | gravitational mass |
| $M_{MAC}$ | macrocosmos mass |
| $M_{MIC}$ | microcosmos mass |
| $M_p$ | Planck mass |
| $M_{UNI}$ | universe mass |
| $p$ | momentum |
| $p_0$ | proper momentum |
| $r_0$ | eigen-particle radius |
| $r_g$ | graviton radius |
| $R_0$ | eigen-universe radius |
| $R_c$ | space-time continuum radius ($R_c = ct_c$) |
| $R_G$ | gravitational radius |
| $R_S$ | Schwarzschild radius |
| $R_{UNI}$ | universe radius |
| $s$ | relativistic space-time interval |
| $s_0$ | quantum space-time interval |
| $t$ | relativistic time |
| $t_0$ | relativistic proper time |
| $t_c$ | age of the space-time continuum |
| $t_r$ | $r_0/c$ |
| $t_{UNI}$ | age of the universe |
| $T_0$ | $1/v_0$ |
| $T_{MAC}$ | macrocosmos time |
| $T_{MIC}$ | microcosmos time |
| $T_p$ | Planck time |
| $T_r$ | $R_0/c$ |
| $v_p$ | particle velocity |
| $x$ | first space co-ordinate |
| $y$ | second space co-ordinate |
| $z$ | third space co-ordinate |
| $\alpha$ | fine structure constant |
| $\lambda_c$ | Compton wavelength |
| $v$ | frequency |
| $v_0$ | proper frequency of a particle |
| $v_g$ | graviton proper frequency |
| $v_H$ | Hubble frequency |
| $N$ | eigen-universe frequency |
| $N_0$ | proper eigen-universe frequency |
| $\tau_0$ | quantum-relativistic proper time ($s_0/c$) |
| $\Psi$ | wave function |
| $\Psi^*$ | complex conjugated wave function |
| $\Psi^{-1}$ | inverse of the wave function |

**Vectors, four-vectors and four-operators**

| | |
|---|---|
| $\vec{x}$ | three-dimensional vector position $(x,y,z)$ or $(x_1, x_2, x_3)$ |
| $s$ | four-dimensional vector position $(ct, ix, iy, iz)$ or $(s_1, s_2, s_3, s_4)$ |
| $[s]$ | four-operator position $(c\hat{t}, i\hat{x}, i\hat{y}, i\hat{z})$ or $(\hat{s}_1, \hat{s}_2, \hat{s}_3, \hat{s}_4)$ |
| $\vec{p}$ | three-dimensional vector momentum $(p_x, p_y, p_z)$ or $(p_1, p_2, p_3)$ |
| $p$ | four-vector momentum $(E/c, ip_x, ip_y, ip_z)$ or $(p_1, p_2, p_3, p_4)$ |
| $[p]$ | four-operator momentum $(\hat{E}/c, i\hat{p}_x, i\hat{p}_y, i\hat{p}_z)$ o $(\hat{p}_1, \hat{p}_2, \hat{p}_3, \hat{p}_4)$ |
| $\nu$ | four-dimensional vector frequency $(\nu_1, \nu_2, \nu_3, \nu_4)$ |
| $[\nu]$ | four-operator frequency $(\hat{\nu}_1, \hat{\nu}_2, \hat{\nu}_3, \hat{\nu}_4)$ |
| $g$ | four-dimensional vector gravitation $(g_1, g_2, g_3, g_4)$ |
| $[g]$ | four-operator gravitation $(\hat{g}_1, \hat{g}_2, \hat{g}_3, \hat{g}_4)$ |

**Special notations**

| | |
|---|---|
| $\circ$ | dot product |
| $\rightarrow$ | application from vector to operator, in particular: |
| $s \rightarrow [s]$ | the four-vector $s$ is transformed into the four-operator $[s]$ |
| $s_1 \rightarrow \hat{s}_1$ | the $s_1$ component of the four-vector $s$ is transformed in the $\hat{s}_1$ component of the four-operator $[s]$ |